\documentclass[prd,floatfix,nofootinbib,showkeys]{revtex4-2}
\usepackage[utf8]{inputenc}
\usepackage[T1]{fontenc}
\usepackage{amsmath}
\usepackage{amsfonts}
\usepackage{amssymb}
\usepackage{graphicx,epstopdf}
\usepackage{datetime}
\usepackage{hyperref}
\usepackage{xcolor}
\usepackage{hyperref}

\begin{document}
	\title{Calorimetric analysis for long-baseline neutrino experiments}

	\author{K. Gallmeister}
	\author{U. Mosel}
		\email[Contact e-mail: ]{mosel@physik.uni-giessen.de}
	\affiliation{Institut f\"ur Theoretische Physik, Universit\"at Giessen, 35392 Giessen, Germany}

	\date{\today}


\begin{abstract}
	In neutrino long-baseline experiments the energy of the incoming neutrino must be reconstructed from observations of the final state. We first discuss the problems that arise for energy conservation during the final-state interactions in a momentum-dependent potential. We then show, for the example of the Deep Underground Neutrino Experiment, that the presence of such a potential is necessarily connected with a large uncertainty in the reconstructed energy. This same uncertainty also affects the determination of energy- and four-momentum-transfers in neutrino-nucleus reactions. We analyze the origins of these uncertainties using transport theory for the description of the evolution of the final state of the reaction. We show that the spectral functions of target nuclei play an essential role not only for the initial neutrino-nucleus interaction but also for the final state evolution.

\end{abstract}

\maketitle

\section{Introduction}

The extraction of neutrino mixing parameters and a possibly CP-violating phase from measurements at long-baseline experiments requires the knowledge of the neutrino energy on an event-by-event basis. Since the incoming neutrinos have a rather broad distribution of energies, the incoming neutrino energy in a given event has to be reconstructed from final-state observables \cite{,Ankowski:2015kya,Mosel:2016cwa}. 

For the lower-energy experiments such as T2K and MicroBooNE this is usually being done by the so-called kinematical method in which measurements on the outgoing lepton are deemed to be sufficient to reconstruct the incoming energy. This method requires a separation of 'true' quasielastic (QE) scattering, which cannot directly be measured, from pion-production and absorption processes. Because of the limited number of open reaction channels at the fairly low energies of the mentioned experiments this separation is under reasonable control if pion production and absorption and 2p2h excitations are understood and tested. In higher-energy experiments such as the Deep Underground Neutrino Experiments (DUNE) with many pions and nucleons in the final state this method works only if very specific exclusive final states are used \cite{Mosel:2013fxa}.

On the other hand, at higher-energy experiments such as DUNE, where many reaction channels are open, the calorimetric method is used in which, in principle, one measures the energies of all the final-state particles and - by energy conservation - thus determines the incoming neutrino energy \cite{Ankowski:2015jya}. In practice, however, there are experimental limitations to this method. Some particles (such as, e.g., neutrons or the target remnants) may evade detection and, in addition, detectors have kinematical detection thresholds. In a very recent publication Coyle et al.\ have shown that such limitations and the shortcomings of generators can affect the extracted fundamental neutrino properties \cite{Coyle:2025xjk}.

In the following we assume that the detector is perfect and study how well the calorimetric method works in determining the true incoming neutrino energy. For such a study one  has to follow the complete reaction from its initial state, consisting of a target nucleus in its bound ground state and an incoming neutrino, to all the final-state particles. The final state is not only determined by the very first, initial neutrino-nucleon interaction (such as QE scattering, resonance excitation, meson exchange coupling or Deep Inelastic Scattering (DIS)) but also by the final state interactions (FSI) of hadrons produced in this initial process.

These FSI have two effects: first, the produced hadrons have to propagate through a nuclear potential before they become free and reach the detector. Second, on their way out, they undergo one-body (e.g. resonance decays),elastic and inelastic  two-body (NN or $\pi$N) and three-body (e.g. in pion absorption) collisions. 

 It is the purpose of this letter to point out some consequences of these in-medium collisions and the in-medium propagation. In particular, we investigate their effects on the calorimetric analysis of the reaction.

\section{Model}
For the present analysis we use the theory framework and generator GiBUU \cite{gibuu} which is based on the quantum-kinetic Kadanoff-Baym non-equilibrium transport theory \cite{Kad-Baym:1962}. The underlying theory is described in \cite{Buss:2011mx} together with details of the numerical implementation and with applications to various nuclear reactions, going all the way from A + A to $\nu$ + A. Further details on the preparation of the ground state as well on the treatment of various background channels can be found in \cite{Gallmeister:2016dnq,Mosel:2023zek}. Here we just summarize some properties that are relevant to the final state propagation and interactions. 

GiBUU uses a consistent nuclear potential both for the ground state and for the final-state propagation of produced hadrons. The potential is position($\vec{r}$)- and momentum($p$)-dependent. In its $p$-dependence it closely mimicks that extracted from pA collisions \cite{Al-Khalili:1990vta,Cooper:1993nx}.  It has been realized early on that the nuclear potential is essential, not only for exclusive final states, but also for inclusive cross sections \cite{Rosenfelder:1978qt,Leitner:2006ww,Ankowski:2014yfa}. This same potential must then also be present during the final-state propagation of baryons.

The $r$-dependence requires the numerical integration of the final-state hadron's trajectory in the nuclear potential. Simple recipes such as a mean-free-path approach, often used in generators \cite{GENIE:2021npt}, can, therefore, no longer be used since it assumes a straight-line trajectory of the particles between collisions. 

The $p$-dependence of the nuclear potential introduces additional complications. It changes, first, the overall cross sections because the initial flux and the final state phase space are changed; this is taken into account by using a result from Pandharipande and Pieper \cite{Pandharipande:1992zz}. In addition, it also complicates the simultaneous energy- and momentum conservation in each collision. This is easy to see for
the simple, nonrelativistic case of a $2 \rightarrow 2$ collision with equal-mass partners. In the cm system one has $\vec{p}_1 = - \vec{p}_2$ so that momentum-conservation is easily obtained. Energy conservation is, however, more complicated. Here one has 
\begin{equation}
	E_{\rm in} = 2\,\left(\frac{{p^\prime}^2}{2m_N} + V(p^\prime) \right)
\end{equation}
where $E_{\rm in}$ is the incoming cm energy in the initial state of the 2-body collision, $p^\prime$ the momentum of the outgoing nucleons and $V(p^\prime)$ their potential. In order to fulfill this equation (in its relativistic extension) GiBUU performs an iterative, numerical solution for each collision \cite{Buss:2011mx} which becomes even more complicated for three-body collisions in, e.g., pion absorption.
 
The final state of this collision may be Pauli-blocked if one of the outgoing momenta lies below the Fermi-surface. An additional effect of the presence of the potential is the binding energy. This comes into play not only when the nucleons finally leave the target but also when one of the collision partners which was initially inside the Fermi-sea is raised to energies above the Fermi-surface while still being inside the nuclear potential. The binding energy is determined by the energy-distribution of bound states in the target. The same is true for the overall momentum conservation where the momentum-distribution of the target nucleons comes into play. The bound-state spectral function thus enters not only in the first neutrino-nucleon collision, but also in the following final-state NN collisions.

In the following analysis we use GiBUU to produce final state events in reactions of neutrinos with $^{40}$Ar nuclei, using the DUNE flux for the incoming neutrino energy distribution. GiBUU delivers the four-vectors of all final state hadrons and is known to describe actual data both at the lower energies (MicroBooNE, T2K) and the higher energies (MINERvA) quite well \cite{Mosel:2023zek}. The final state distributions are thus assumed to be quite realistic and close to "nature". As usually done in generators we use the impulse approximation and assume the validity of the so-called frozen configuration that keeps the target density and potential constant during the final state evolution. Also, as usual in experimental analyses, the energy and the recoil momentum of the target remnant are neglected. We then analyze the events so generated by the calorimetric method. 

\section{Calorimetric analysis}

For the DUNE we find that up to about 20 nucleons are unbound in the final state, with additional up to 6 nucleons being excited, but still bound in the potential. The nucleons still bound in the target remnant carry an excitation energy of about 6 MeV each \cite{Ferrari:1995cq}, typical of a compound nucleus. The decay of that compound nucleus will - at later times - also contribute to baryon spectra but with baryon energies well below neutrino detector thresholds. 

For the calorimetric analyses we start the discussion with the seemingly natural assumption that only the kinetic energies of the outgoing free nucleons and mesons 
determine the energy-balance for the reaction. The underlying assumption is that for incoming neutrino energies in the GeV region, such as for the DUNE, potential effects of the order of a few 10's of MeV can be neglected. The missing energy, defined as the difference between the energy transfer and the final-state energies, is then given by
\begin{equation}    \label{Emiss1}
	E_{\rm miss}\simeq\nu-\sum_{i=1}^{N} T_i-\sum_{j=1}^{M}E_j\ .
\end{equation}
\begin{figure}
	\centering
	\includegraphics[width=0.7\linewidth]{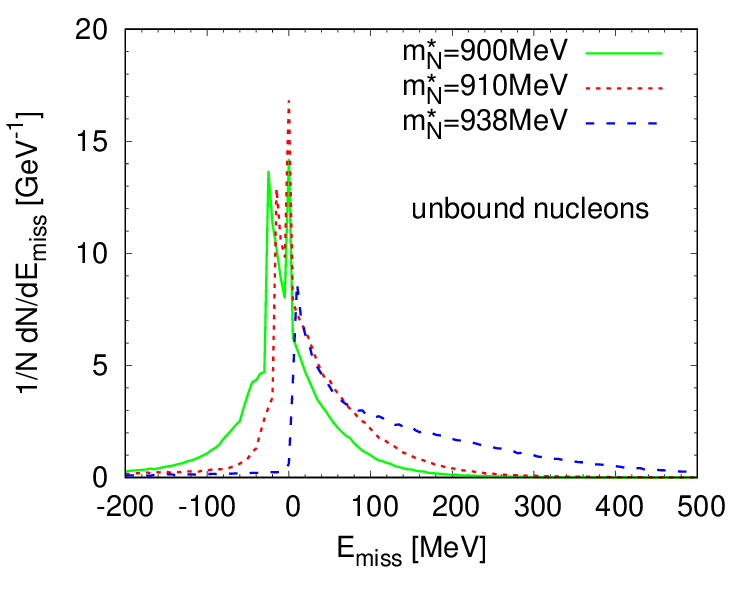}
	\caption{(color online) Missing energy distribution for the DUNE flux on an $^{40}$Ar target for three different in-medium masses $m_N^*$. Only free final-state nucleons and mesons are used for the determination of the missing energy. The long-dashed curve gives the result of an analysis using Eq.\ (\ref{Emiss1}), the short-dashed and the solid curves show the results for average binding energies of 28 MeV and 38 MeV, respectively, in Eq.\ (\ref{Emiss2}).}
	\label{fig1}
\end{figure}
 Here $\nu$ is the energy transfer, $N$ stands for the number of observed, free nucleons and $M$ is the number of final-state mesons; their free energies are denoted by $E_j$, the kinetic energies of the outgoing nucleons are given by $T_i$. 
 
 The missing energy distribution obtained in this way from the events generated in a GiBUU calculation ("nature") is shown by the long-dashed (blue) curve in Fig.\ \ref{fig1}. It exhibits a broad tail towards positive missing energies.  Naively, one could conclude from this result that the generator used to generate the events violates energy conservation. However, this tail reflects the fact that the target nucleons in GiBUU are bound, with a separation energy of about 8 MeV for nucleons at the Fermi-surface and a maximum binding of about 70 MeV. Obviously, these binding energy effects become the more important the higher the multiplicity of outgoing nucleons (up to 20)  is.

As an improvement of this simple approximation we, therefore, now perform the same analysis but with including explicitly an average binding energy for the outgoing nucleons. The missing energy is now given as
\begin{eqnarray}   \label{Emiss2}
	E_{\rm miss} &\simeq& \nu-\sum_{i=1}^NT_i-N B_N - \sum_{j=1}^{M}E_j\nonumber \\
	&=& \nu -\sum_{i=1}^NE_i-\sum_{j=1}^{M}E_j\ + N m_N^* .
\end{eqnarray}		
 $B_N$ is an average binding energy of the outgoing nucleon, in experimental analyses  often taken to be $B_N \approx 30$ MeV. In the second line of Eq.\ (\ref{Emiss2}) we assume that the binding potential is scalar and therefore the in-medium mass of the nucleon is given by $m_N^* = m_N - B_N$ where $m_N$ is the nucleon mass taken to be 938 MeV. The result of a calculation of the missing energy thus defined is shown by the short-dashed (red) and the solid (green) curves in Fig.\ \ref{fig1}. While the large, positive-energy tail of the distribution has now shrunk significantly a tail towards negative missing energies develops thus making the  distribution more symmetric around the value $E_{\rm miss} = 0$. The width of these distributions then determines the implicit uncertainty in the determination of the incoming neutrino energy by the calorimetric method, even when the detectors are perfect. Coyle et al. have recently shown that two other generators, GENIE and NuWro, give narrower distributions for the missing energy \cite{Coyle:2025xjk}. This seeming advantage is just a consequence of the neglect of a final state potential in these generators.
 
Up to this point GiBUU has been used only to generate the events ("mock data") but not for their analysis; the missing mass determination is based only on unbound final states and is independent of this specific generator. We now use GiBUU also for a better understanding of the results just obtained. The energy balance of the $\nu$A reaction is affected not only by the final-state free hadrons, but also by hadrons that are produced during the FSI but never make it out of the target.
The latter number is experimentally not accessible, but GiBUU delivers some insight into their effects. The missing energy is now given by
\begin{align}   \label{Emiss3}
	E_{\rm miss}=\nu+\sum_{i=N+1}^{N'}(E'_i-E_i)+\sum_{i=1}^{N}E'_i-\sum_{i=1}^{N}E_i-\sum_{j=1}^{M}E_j\ .
\end{align}
Here $N$ stands for the number of observed nucleons and $N'$ for that of all nucleons involved in the FSI, $E'$ indicates the energy of the hit nucleon bound in the nucleus, while $E$ stands for the energy of the final (free) nucleon (resp.~meson). Even when the
energy of the scattered nucleon is not large enough for the nucleon to escape, it is definitively larger than the energy of the original nucleon in the Fermi sea. Thus, the first sum, the contribution of collisions with participating nucleons not making it out of the nucleus, is smaller than zero. A detailed analysis shows that the number of nucleons involved in collisions, but not making it out of the nucleus, is about one half of the number of free, observed nucleons.

\begin{figure}
	\hspace*{\fill}%
	\includegraphics[width=0.7 \linewidth,clip=true]{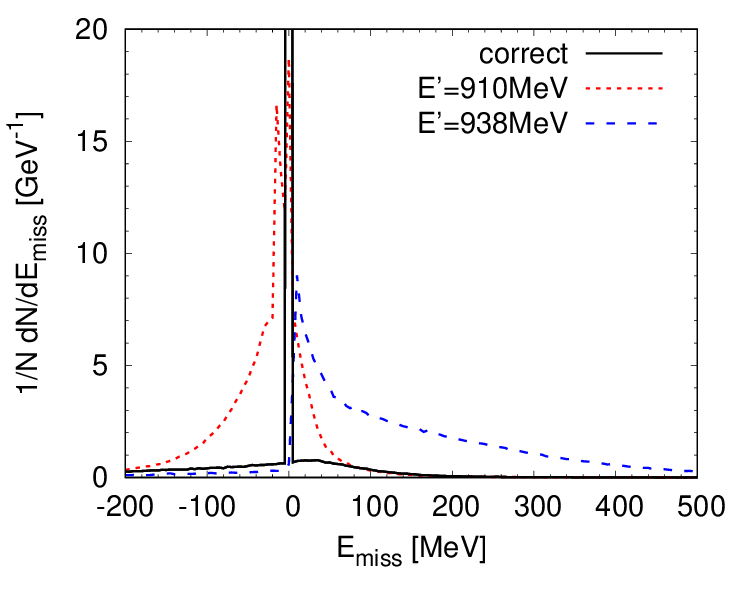}
	\hspace*{\fill}%
	\caption{Missing mass distribution for a $\nu$ + Ar collision with the DUNE incoming neutrino flux. All nucleons involved in the collision are included. The curves labelled '938MeV' and '910MeV' are the result of using Eq.\ \ref{Emiss3}. } 
	\label{fig2}
\end{figure}

The results of this analysis are shown in Fig.\ \ref{fig2} by the solid (black, 'correct') curve. This illustrates that energy conservation is nearly perfect in GiBUU when the nucleon's energy distribution in the target is described correctly. The very low broad pedestal distribution is due to DIS events obtained with the PYTHIA event generator. The other two curves in Fig.\ \ref{fig2} give the results obtained when replacing the energies $E'$ in Eq.\ (\ref{Emiss3}) by the average values given in the figure. While the long-dashed curves (E' = 938 MeV) look very similar in both figures the curve with an average Fermi-sea mass of 910 MeV (dotted curves) is now shifted towards negative energies.

\vspace*{\fill}

\section{Summary and Conclusion}
The calorimetric energy reconstruction in a neutrino long-baseline experiment faces some obstacles connected with the in-medium properties of nucleons in the nuclear target that affect the final-state collisions. In particular we have shown that a simple calorimetric energy reconstruction, based on the observed final state hadrons, with or without binding energy corrections, leads to a significant uncertainty in the missing energy distribution and thus the reconstructed neutrino energy. 

In an analysis of actual data this problem appears since there is no direct experimental handle neither on the initial energies of the target nucleons nor on the intermediate nuclear collisions. Theoretically it can partly be overcome only by generators that contain a consistent in-medium potential and proper energy-momentum conservation in final state collisions inside the potential range. The generators must also be able to deal with the binding energies in  collisions in which the participating nucleons do not make it out of the nucleus but remain bound. The nucleon's spectral functions thus do not only play a role in the initial neutrino-nucleus reaction but also in the final-state collisions thus affecting the energy reconstruction.

The problems investigated here also affect the experimental reconstruction of the energy transfer and thus also of the four-momentum transfer $Q^2$ in neutrino long-baseline experiments. In experimental analyses often an 'available energy' is obtained by measuring energies of some final-state particles. The identification of this 'available energy'  with the true energy transfer is bound to be plagued by the same uncertainties as just discussed for the overall energy conservation.  

We finally note that also the neglect of target remnant effects beyond the single-particle model (excitation, momentum) as well as the assumption of a frozen configuration deserve further study.

\begin{acknowledgments}
The authors are grateful to N. Coyle whose neutrino oscillation analyses motivated this present Letter.
\end{acknowledgments}

\bibliography{nuclear.bib}
 
\end{document}